\documentclass[showpacs,amsmath,amssymb,aps,prl,preprint,superscriptaddress]{revtex4-1}

\usepackage{graphicx}
\usepackage{bm}
\usepackage{epstopdf}

\begin{document}

\title{Excited-State Structure Modifications due to Molecular Substituents and Exciton Scattering in Conjugated Molecules}

\author{Hao Li}
\affiliation{Theoretical Division, Center for Nonlinear Studies, Los Alamos National Laboratory, Los Alamos, NM 87545}

\author {Michael J. Catanzaro}
\affiliation{Department of Mathematics, Wayne State University, 656 W. Kirby, Detroit, MI 48202}

\author {Sergei Tretiak}
\email{serg@lanl.gov}
\affiliation{Theoretical Division, Center for Nonlinear Studies, Los Alamos National Laboratory, Los Alamos, NM 87545}
\affiliation{Center for Integrated Nanotechnologies, Los Alamos National Laboratory, Los Alamos, NM 87545}

\author {Vladimir Y. Chernyak}
\email{chernyak@chem.wayne.edu}
\affiliation{Department of Chemistry, Wayne State University, 5101 Cass Avenue, Detroit, MI 48202}
\date{January 28, 2014}




\begin{abstract}
Attachment of chemical substituents (such as polar moieties) constitutes an efficient and convenient way to modify physical and chemical properties of conjugated polymers and oligomers. Associated modifications in the molecular electronic states can be comprehensively described by examining scattering of excitons in the polymer's backbone at the scattering center representing the chemical substituent. Here, we implement effective tight-binding models as a tool to examine the analytical properties of the exciton scattering matrices in semi-infinite polymer chains with substitutions. We demonstrate that chemical interactions between the substitution and attached polymer is adequately described by the analytical properties of the scattering matrices. In particular, resonant and bound electronic excitations are expressed via the positions of zeros and poles of the scattering amplitude, analytically continued to complex values of exciton quasimomenta. We exemplify the formulated concepts by analyzing excited states in conjugated phenylacetylenes substituted by perylene.
\end{abstract}

\maketitle

\begin{figure}[htp]
 \begin{center}
 \includegraphics[width=5in]{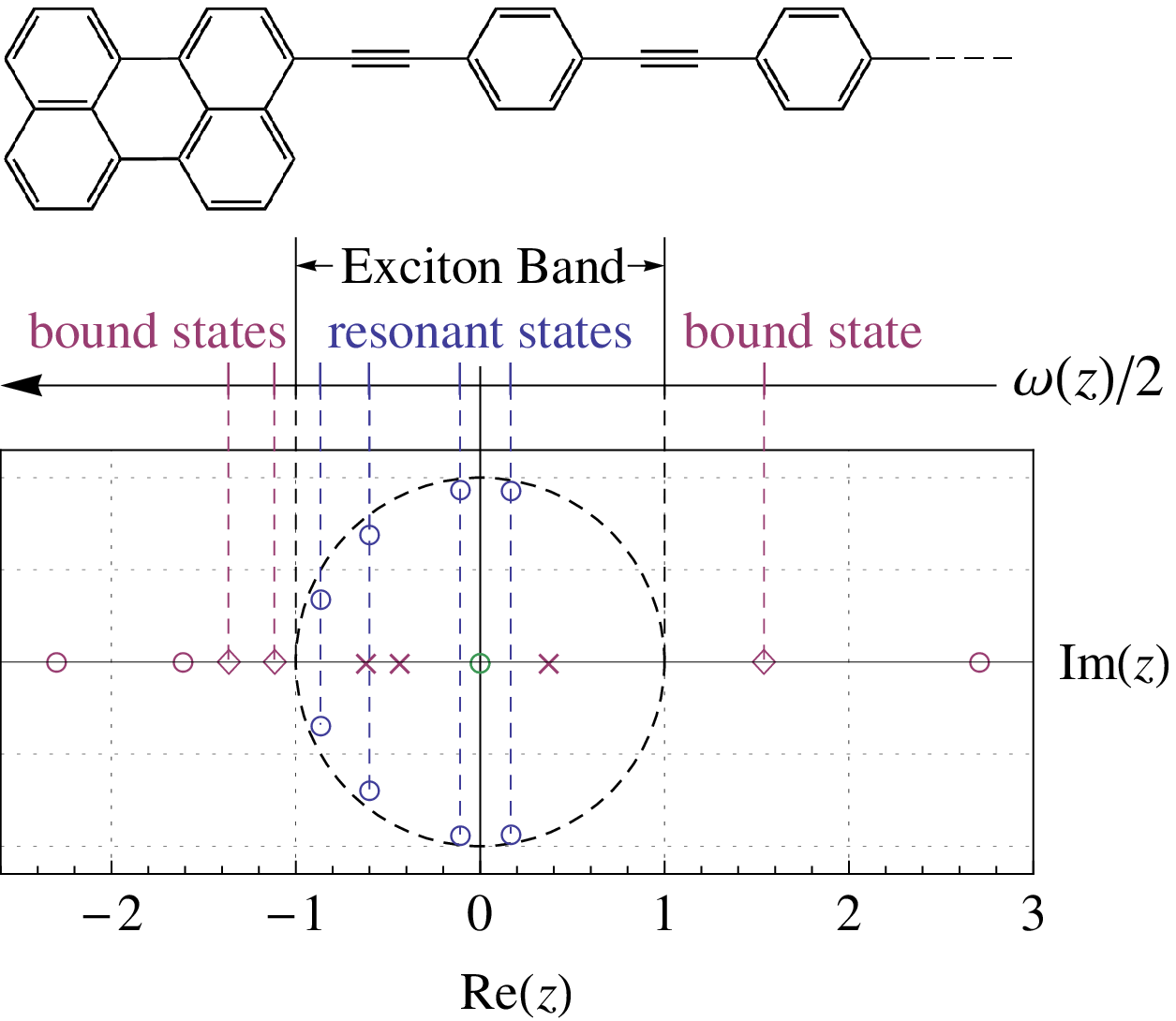}
 \end{center}
\end{figure}


\newpage

The concept of scattering has proven to play an important role in the description of a variety of phenomena in quantum mechanics, condensed matter and quantum field theory, functional analysis, and chemistry. Scattering matrices/operators contain detailed information on fundamental interactions that control the system dynamics, and are available from experimental measurements. In high-energy physics the differential cross-sections, directly related to scattering matrices, are measured and further interpreted using quantum field theory with the ultimate goal of revealing the fundamental interactions. In condensed matter theory, the scattering matrices provide a useful tool of re-summing the short range interaction effects, e.g., in the superfluidity theory in the low-density case, where a perturbation theory can be formulated in terms of the particle-particle scattering matrix that contains all necessary information on the particle-particle interactions \cite{Landau:superfluidity,superfluidity,Kokkelmans:RSuperfluidity}. The celebrated Fermi-liquid theory allows interpretation of complex dynamics of a strongly interacting system to be interpreted in terms of quasi-particle spectra and scattering matrices \cite{Landau:Fermi-liquid1957,Fermi-liquid}. Chemical reactions in gas phase can be conveniently formulated and interpreted as scattering processes with the asymptotic states represented by the reactants and products \cite{Zhang:GasPhaseReactions,Rebentrost:JCP1977}. In functional analysis, scattering theory can be viewed as a tool of analyzing continuous spectra of unbounded (often differential) operators, in particular the projection measures involved in spectral decompositions \cite{FunctionalAnalysis}.

Our recent work \cite{ES-naturephys,ES-prl,ES1,ES2,ES3,ES-dipole,ES-DA,ES-NAO} has demonstrated that the quasiparticle picture, coined Exciton Scattering (ES) approach, provides a simple and clear insight into the excited-state electronic structure in complex conjugated macro-molecules \cite{Heeger:NobelLecture,Heeger:RevModPhys1988,BredasJL:Excesc,heeger2010semiconducting,DreuwA:Sinaim}. It allows excited electronic states to be studied in terms of the exciton spectra in infinite polymers and scattering matrices to be associated with molecular vertices, i.e., termini, joints and branching centers \cite{Mullen:ChemRev1999,KopelmanR:Speeel,PercecV:Selshd,Bunz:AdvPolymSci05,ahn:jcp06,Moore:awie06,HaleyMM:ChemRev06}. The exciton scattering properties of molecular vertices can be further described by tight-binding or equivalently lattice models  \cite{ES-lattice}. This extends the exciton scattering concept to the case of imperfect molecular geometries aimed at deriving the exciton-phonon Hamiltonian, thus mapping the problem of incoherent energy transfer in branched conjugated structures onto a much simpler (although still complex) counterpart of incoherent motion of Frenkel-type excitons \cite{MinamiT:FreHdn,fidder:7880}. The scattering framework in the conjugated molecular systems is different from typical cases usually studied in quantum mechanics, due to discrete, rather than continuous translational symmetry of the asymptotic states. In particular, integer topological invariants, namely winding numbers/topological charges, can be associated with the scattering centers \cite{ES-symmetric}. It is well established in quantum mechanics that analytical properties of scattering matrices, more specifically their analytic continuations, provide detailed and important information on the underlying potentials, including bound and metastable states. Therefore, analytical properties of exciton scattering on molecular vertices are expected to provide adequate information of how electronic properties of supramolecular conjugated structure are affected by local chemical substituents.

\begin{figure}[th]
  \includegraphics[scale=0.8, trim={0 8cm 0 1cm},clip]{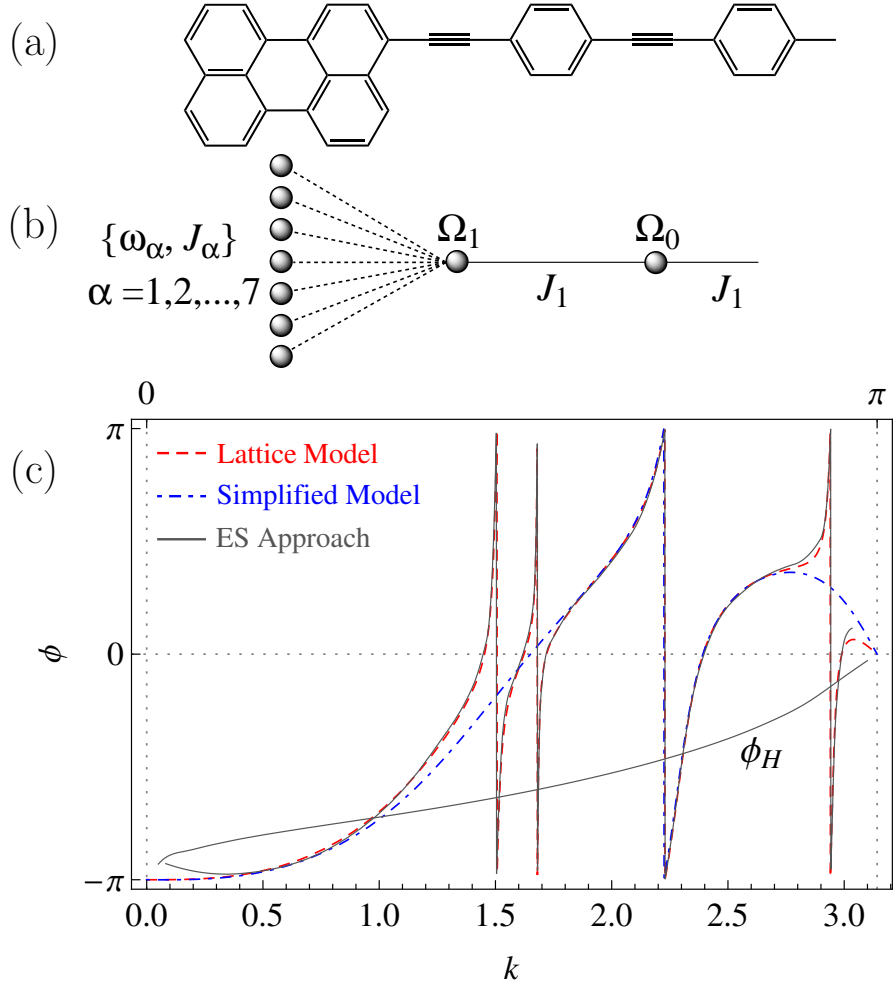}
\caption{
\label{fig:perylene-linear}
 (a) Perylene terminated phenylacetylene (PA) linear molecule; (b) The tight-binding model describing the exciton scattering at the terminus; (c) The exciton reflection phase of the terminus obtained from the ES approach (black solid curve), which is approximated by the tight-binding model showed in (b) (red dashed curve), and by the simplified model (blue dot dashed curve), where three sharp resonances have been removed. Reflection phase of unmodified terminus $\phi_H$ is showed for comparison.
}
\end{figure}

In this manuscript, we apply tight-binding models as a tool to study analytical properties of scattering matrices, providing insights into the excited-state electronic structure of molecular substituents commonly present in conjugated systems. As an example, Figure~\ref{fig:perylene-linear}c displays the reflection phase of perylene attached to a semi-infinite phenylacetylene (PA) chain (see Fig.~\ref{fig:perylene-linear}a), retrieved from quantum chemistry calculations \cite{ES-DA}. Compared to that in an unsubstituted chain ($\phi_H$), the plot shows highly non-trivial dependence of the phase on the exciton quasimomentum $k$, including $4$ resonant features, that are not easy to interpret. Here we show how these features in the reflection phase can be directly ascribed to resonant and bound excited states brought by the substituent.

We start with introducing the simplest nearest-neighbor hopping lattice model, where the linear segments of a branched conjugated structure are represented by linear chains (graphs) with the same on-site energy $\bar{\Omega}_{0}$ and the same hopping constant $\bar{J}_{1}$ between the nearest neighbors. A molecular vertex is represented by a complete graph, i.e., by a set of fully interconnected sites, with arbitrary on-site energies and hopping constants between any pair of sites allowed. The chemical connection between a molecular vertex and the attached linear segment is thus described by introducing the hopping constants between the first lattice site (whose on-site energy is also modified) of the linear segment and any lattice site that belongs to the vertex (e.g., see Fig.~\ref{fig:perylene-linear}b). Here we consider the case of 
chemical substitution on molecular terminus only (e.g., perylene in Fig.~\ref{fig:perylene-linear}a); the general case of an arbitrary degree molecular vertex will be analyzed elsewhere. Since the lattice sites representing a vertex form a basis set for the vertex tight-binding Hamiltonian, without loss of generality we can assume the latter to be diagonal. Therefore, a molecular terminus is described by a set $\{\omega_{\alpha}|\alpha=1,\ldots,n\}$ of the on-site energies, a vector $\bm{J}=(J_{\alpha}|\alpha=1,\ldots,n)$ of the hopping constants between the vertex sites and the first linear segment site, and the modified on-site energy $\Omega_{1}$ of the latter, as shown in Fig.~\ref{fig:perylene-linear}b. Measuring the energy in the unit of $\bar{J}_1$ and choosing the zero energy level at the middle of the exciton band, without loss of generality we can set $\bar{\Omega}_{0}=0$ and $\bar{J}_{1}=1$.

Within the described lattice model, the exciton wavefunction on a semi-infinite chain is given by the sets $\bm{\Psi}=(\Psi_{\alpha}|\alpha=1,\ldots,n)$ and $\bm{\psi}=(\psi_{j}|j=0,1,\ldots)$ of its values on the terminus and the chain respectively. Introducing the multiplicative variable $z=e^{ik}$ that describes the quasimomentum $k$, we represent the wavefunction on the chain as a superposition of incoming and outgoing waves
\begin{eqnarray}
\label{eq:psi-chain} \psi_{j}= z^{-j}+ r(z)z^{j}, \;\;\; j=0,1,\ldots
\end{eqnarray}
with $r(z)$ being the quasimomentum-dependent reflection coefficient at $j=0$. The eigenmode equation adopts a form
\begin{eqnarray}
\label{eq:psi-equations} \psi_{j-1}+\psi_{j+1}&=& \omega\psi_{j}, \;\;\; j=1,2,\ldots \nonumber \\ \omega_{\alpha}\Psi_{\alpha}+J_{\alpha}\psi_{0}&=& \omega\Psi_{\alpha}, \;\;\; \alpha= 1,\ldots,n \nonumber \\ \sum_{\alpha}J_{\alpha}\Psi_{\alpha}+ \psi_{1}&=& (\omega-\Omega_{1})\psi_{0}.
\end{eqnarray}
Upon substituting Eq.~(\ref{eq:psi-chain}) into Eq.~(\ref{eq:psi-equations}) we can easily solve the system of linear equations and obtain the following expressions for the exciton spectrum
\begin{eqnarray}
\label{eq:spectrum} \omega(z)= z+z^{-1}
\end{eqnarray}
and reflection coefficient
\begin{eqnarray}
\label{eq:reflection-coeff} r(z)= -\frac{z - \Omega_{1}-F(\omega)}{z^{-1} - \Omega_{1}-F(\omega)}, \; F(\omega)=\sum_{\alpha}\frac{J_{\alpha}^{2}}{\omega-\omega_{\alpha}},
\end{eqnarray}
both being represented by meromorphic functions of $z$ in the complex plane. This implies that $r(z)$ can be also interpreted as a meromorphic function on the projective space $\mathbb{C}P^{1}$ or a holomorphic function $r:\mathbb{C}P^{1}\to \mathbb{C}P^{1}$. These two interpretations originate from viewing $\mathbb{C}P^{1}$ as a compact complex analytical manifold (of complex dimension $1$) topologically equivalent to a sphere $\mathbb{C}P^{1} \cong S^{2}$ with the complex structure induced from complex plane $\mathbb{C}$ via stereographic projection $\mathbb{C}\hookrightarrow \mathbb{C}P^{1}$, i.e., $\mathbb{C}P^{1}$ is obtained from $\mathbb{C}$ by adding the infinite point.

The reflection coefficient satisfies obvious relations
\begin{eqnarray}
\label{eq:r-symmetry} r(z^{-1})=(r(z))^{-1}, \;\;\; r(z^{*})=(r(z))^{*}
\end{eqnarray}
that reflect unitarity of quantum mechanics combined with time-reversal symmetry.

A direct inspection of Eq.~(\ref{eq:reflection-coeff}) shows that the reflection coefficient can be represented in a form $r(z)=zP_{2n+1}(z)/Q_{2n+1}(z)$, with $P_{2n+1}$ and $Q_{2n+1}$ being polynomials of degree $(2n+1)$, which means that $r(z)$, as a meromorphic function in $\mathbb{C}P^{1}$, has $2(n+1)$ zeros and $2(n+1)$ poles, or equivalently that the map $r:\mathbb{C}P^{1} \to \mathbb{C}P^{1}$ has degree $n_{{\rm A}}(r)=2(n+1)$, hereafter referred to as the analytical index. 
Due to the symmetry relations [Eq.~(\ref{eq:r-symmetry})], the roots of $Q_{2n+1}$ are inverse to the roots of $P_{2n+1}$ with the roots of each polynomial being either real or coming in mutually complex conjugated pairs. This implies that the positions of the roots, of say $P_{2n+1}$, fix $r(z)$ up to a multiplicative factor, the latter being fixed by the condition $r(1)=-1$ [note that also $r(-1)=-1$] that follows from Eq.~(\ref{eq:reflection-coeff}) [note that generally Eq.~(\ref{eq:r-symmetry}) also implies $r(1)=\pm 1$, as well as $r(-1)=\pm 1$]. Therefore $r(z)$ can be represented in a form
\begin{eqnarray}
\label{eq:reflection-coeff-roots} r(z)= z\prod_{j=1}^{m}\frac{w_{j}z-1}{z-w_{j}}\prod_{j=m+1}^{2n+1}\frac{z-z_{j}}{z_{j}z-1},
\end{eqnarray}
where $\{z_{j}\}$ and $\{w_{j}\}$ are the zeros of the polynomials $P_{2n+1}$ and $Q_{2n+1}$, respectively, located inside the circle $|z|=1$, hereafter referred to as the circle.

We now tie the analytic properties described above together with the underlying chemistry
in terms of bound states.
According to quantum-mechanical scattering theory and Eq.~(\ref{eq:psi-chain}), the bound states in a semi-infinite chain correspond to the poles of $r(z)$ located inside the circle, or equivalently the zeros located outside, so that the energies of the bound states are given by $\omega(w_{j})$, which implies that $w_{j}$ should be real for all $j=1,\ldots,m$. In our earlier work (unpublished result), we have introduced the topological index (winding number) $n_{{\rm T}}(r)$, associated with a vertex. A direct calculation yields
\begin{eqnarray}
\label{eq:Q-W} n_{{\rm T}}=\oint_{|z|=1}\frac{dz}{2\pi i}r^{-1}\frac{dr}{dz}= 2n+2-2m.
\end{eqnarray}
Introducing the molecular vertex analytical and topological charges $Q_{{\rm A}}=(n_{{\rm A}}-2)/2$ and $Q_{{\rm T}}=(n_{{\rm T}}-2)/2$, respectively, we conclude that in a finite-length linear molecule with $L$ repeat units, represented by $L$ sites within the nearest neighbor lattice model, the number of exciton states is given by
\begin{eqnarray}
\label{eq:N-analytical} N= L+ Q_{{\rm A}}^{(1)}+ Q_{{\rm A}}^{(2)},
\end{eqnarray}
whereas the number of bound states associated with a vertex $a$ is given by $m^{(a)}= Q_{{\rm A}}^{(a)}- Q_{{\rm T}}^{(a)}$. In a molecule with the lengths of linear segments considerably exceeding the localization lengths of the bound states, we have the number of exciton states inside the exciton band
\begin{eqnarray}
\label{eq:N-topological} N_{0}= L+ Q_{{\rm T}}^{(1)}+ Q_{{\rm T}}^{(2)}.
\end{eqnarray}
This is true since the excitons with energies outside the exciton band are excellently approximated by just the bound states.
Note that Eq.~(\ref{eq:N-topological}) reflects one of the statements of the index theorem for the case of linear molecules presented in our earlier work \cite{ES-symmetric}, and Eq.~(\ref{eq:Q-W}) can be interpreted as a relation between the analytical and topological properties of a molecular vertex: 
the topological index is formed from $(2(n+1)-m)$ positive and $m$ negative contributions associated with the zeros and the poles, respectively, of $r(z)$, located inside the circle.

The analytical index theorem [Eq.~(\ref{eq:N-analytical})] can be also derived in more general terms by noting that the ES equations for a linear molecule can be written in a form
\begin{eqnarray}
\label{eq:ES-eq-linear} \tilde{\Gamma}(z)-1=0, \;\;\; \tilde{\Gamma}(z)=r^{(1)}(z)r^{(2)}(z)z^{2(L-1)},
\end{eqnarray}
so that the number of its solutions is given by the analytical index $n_{{\rm A}}(\tilde{\Gamma}-1)$ (unpublished result). We further observe
\begin{eqnarray}
\label{eq:A-index-derive} n_{{\rm A}}(\tilde{\Gamma}-1)= n_{{\rm A}}(\tilde{\Gamma})= n_{{\rm A}}^{(1)}+n_{{\rm A}}^{(2)}+ 2(L-1),
\end{eqnarray}
and note that there are two unphysical solutions with $z=\pm 1$, whereas each exciton state is represented by a pair of symmetry related solutions with mutually inverse values of $z$. This results in $N= (n_{{\rm A}}(\tilde{\Gamma}-1)/2) -1$, which reproduces Eq.~(\ref{eq:N-analytical}).

It is worth mentioning that shifting the reference point of reflection by $\Delta j$ results in a factor of $z^{2\Delta j}$ in $r(z)$ \cite{ES-symmetric}, and changes of both $n_{{\rm A}}$ and $n_{{\rm T}}$ by $2\Delta j$, which followed by the change of $Q_{{\rm A}}$ and $Q_{{\rm T}}$ by $\Delta j$; the latter is compensated in Eqs.~(\ref{eq:N-analytical}) and (\ref{eq:N-topological}) by the corresponding change of $L$ by $-\Delta j$. For convenience, we define $Q_{{\rm A}}$ and $Q_{{\rm T}}$ using $Q_{{\rm A/T}}=(n_{{\rm A/T}}-2-2\Delta j)/2$, so that the analytical and topological charges (integers) are the intrinsic properties of molecular vertices and independent of the reflection point, and the length $L$ should not change with respect to the choice of the reflection position.

In summary, within a nearest-neighbor hopping lattice model a molecular terminus is described by the reflection coefficient $r(z)$, characterized by its topological charge $Q_{{\rm T}}$ and analytical charge $Q_{{\rm A}} \ge Q_{{\rm T}}$, and represented by a meromorphic function on $\mathbb{C}P^{1}$ in a form given by Eq.~(\ref{eq:reflection-coeff-roots}). It is fully determined by the positions of its $(Q_{{\rm A}}- Q_{{\rm T}})$ poles inside the circle that all lie on the real axis, and $(Q_{{\rm A}}+Q_{{\rm T}}+2)$ zeros, referred to as resonances, inside the circle which either belong to the real axis or come in mutually complex conjugated pairs. Stated equivalently, using a term ``tight-binding model with nearest neighbor hopping'' is equivalent to approximating the scattering coefficient $r(z)$ with a meromorphic function, defined in $\mathbb{C}P^{1}$, since there is a one-to-one correspondence between such meromorphic functions (with the certain simple properties, implied by fundamental quantum mechanical symmetries, and explicitly described earlier in the text) and the sets of parameters of tight-binding models of a certain class, namely, referred to as nearest-neighbor hopping. Therefore, hereafter we will not make a distinction between the terms ``tight-binding model with nearest neighbor hopping'' and ``scattering coefficients represented by meromorphic functions in $\mathbb{C}P^{1}$''. Having said that, we would like to note that dealing with a tight-binding model as a tool has certain advantages by providing a simple and intuitive physical insight, as well as opening the way to efficiently account for exciton-phonon interactions, as proposed in our earlier work \cite{ES-lattice}.

An interesting situation occurs when a pair $(z_{j},z_{j}^{*})$ of resonances lies close to the circle. Introducing a natural notation $z_{j}=(1-\delta_{j})e^{ik_{j}}$ with $\delta_{j} \ll 1$, we can represent the contribution of the above pair of resonances to the reflection coefficient [Eq.~(\ref{eq:reflection-coeff})] in a form
\begin{eqnarray}
\label{eq:r-resonance} r_{j}(z)\approx \frac{z-(1-\delta_{j})e^{ik_{j}}}{z-(1+\delta_{j})e^{ik_{j}}} \frac{z-(1-\delta_{j})e^{-ik_{j}}}{z-(1+\delta_{j})e^{-ik_{j}}},
\end{eqnarray}
where we have applied the approximations $(1-\delta_{j})^{-1}\approx 1+\delta_{j}$ and $(1-\delta_{j})^{2}\approx 1$ (which is irrelevant to the phase factor so that can be ignored). 
Except for the narrow regions in the quasimomentum space of the width $\Delta k_{j}\sim \delta_{j}$ around $k=\pm k_{j}$ we have $r_{j}\approx 1$, i.e., it does not contribute to the reflection coefficient, whereas in the above regions $r(z)$ shows resonant behavior, where the scattering phase $\phi=-i \ln r$ acquires a contribution of $2\pi$ over a narrow region $\Delta k_{j}\sim \delta_{j}$. In a simplified scenario where $\Omega_1=0$, $n=1$ and $m=0$, one can find that $\delta_1\approx J_1^2/2$, which indicates that the sharp resonant feature is attributed to the weak coupling between the terminal site and the chain. These resonances, hereafter referred to as phase kinks, correspond to the resonant states, i.e., excited states on the substituent, weakly coupled to the exciton band in the polymer chain.

In the aforementioned tight-binding model, the $m$ lattice sites, representing the graph of the terminus, related to the bound states simultaneously result in $m$ poles of $r(z)$ outside of the circle, i.e., $m$ zeros located inside the circle. Taking into account a ``trivial" zero of $r(z)$ at $z=0$ and a zero on real axis close to $z=0$, which is attributed to the small value of $\Omega_1$, there are $2(n-m)$ zeros close to and inside of the circle come in mutually complex conjugated pairs which are associated to the $(n-m)$ resonant states within the exciton band. Stated differently, the $(Q_{{\rm A}}- Q_{{\rm T}})$ bound states are directly associated with the poles inside the circle on the real axis, whereas the resonant states correspond to the $Q_{{\rm T}}$ pairs of complex conjugated zeros of $r(z)$ located close to the circle.

To illustrate the above formalism, we calculate electronic excitations in linear phenylacetylene molecules with and without perylene substituent (Fig.~\ref{fig:perylene-linear}a) using standard quantum chemical (QC) methodology. The length of the PA molecules varies from 5 to 35 repeat units with increments of 5 repeat units. As we have shown before \cite{ES-prl,ES-NAO}, any approach for excited-state computation, that can adequately describe exciton properties (including the binding energy and the exciton size), can be used as a reference QC method in the ES approach. Here the ground state geometries have been optimized at the semiempirical Austin Model 1 (AM1) level \cite{cook} using the Gaussian09 package \cite{g09}. We then applied the collective electronic oscillator (CEO) method \cite{ChernyakV:KryatH,TretiakS:Recdam,TretiakS:Denmas}, which is based on the time-dependent Hartree-Fock (TDHF) theory combined with the semiempirical INDO/S (intermediate neglect of differential overlap parameterized for spectroscopy) Hamiltonian \cite{zindo}, to compute the excitation energies, transition dipoles, and transition density matrices. The lowest exciton band has been singled out by inspecting the structures of transition density matrices in real-space \cite{ES2}. The exciton spectrum $k(\omega)$ and the reflection phases of unmodified and perylene-substituted termini have been extracted within the ES approach \cite{ES-DA}. As shown in our previous studies, physically similar results can be expected 
using other model QC techniques for excited state calculations such as time-dependent density functional theory (TDDFT) \cite{ES-NAO}. The ES phase of perylene and the excited-state data provide sufficient input for a corresponding tight-binding model \cite{ES-lattice}: the nearest-neighbor tight-binding parameters in an infinite linear chain, $\bar{\Omega}_0$ and $\bar{J}_1$, can be found from the exciton spectrum $k(\omega)$ and Eq.~(\ref{eq:spectrum}).

We are now in a position to describe the excited-state chemical properties of perylene attached to a PA chain in terms of analytical properties of the reflection phase obtained with a chosen model of QC. The molecular vertex that represents perylene (Fig.~\ref{fig:perylene-linear}a) has the charges $Q_{{\rm A}}=7$ and $Q_{{\rm T}}=4$, which yields $Q_{{\rm A}}- Q_{{\rm T}}=3$ bound states and $Q_{{\rm T}}=4$ resonant states. Using the model depicted in Fig.~\ref{fig:perylene-linear}b, we parameterized the tight-binding graph that represents the perylene terminus by fitting its reflection phase. Specifically, in the nearest-neighbor model of the linear chain, extracted and tabulated quantities include ${\bar \Omega}_0=3.492$~eV, ${\bar J}_1=-0.288$~eV, on-site energy $\Omega_1=3.488$~eV of the first site in the segment, and parameters of the terminus given in Table.~\ref{tab:parameters}. The actual scattering phase, approximated by the above lattice model, can be viewed as the addition of sharp resonant features (Eq.~(\ref{eq:r-resonance})) on top of the simplified model, in which the hopping constants $J_\alpha$ ($\alpha=2,3,6$) have been set to zero, due to the weak couplings between the corresponding states of the terminus and the chain (Fig.~\ref{fig:perylene-linear}b).

\begin{figure}[th]
  \includegraphics[scale=0.6, trim={0 10cm 0 0},clip]{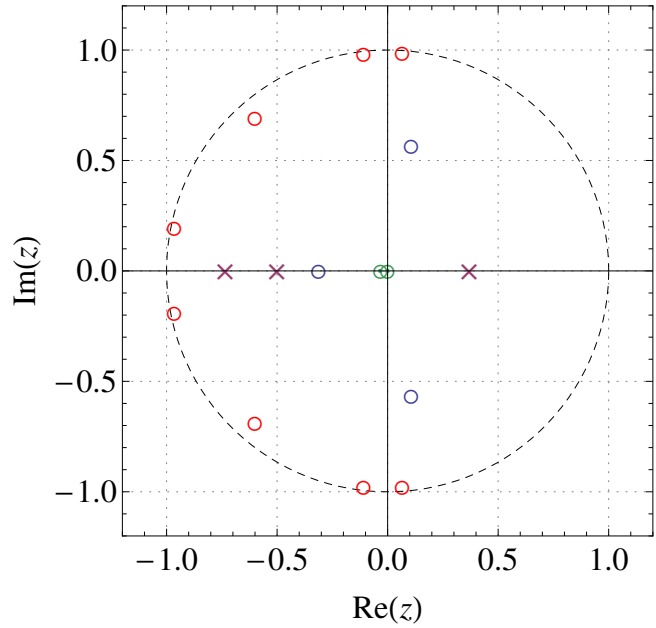}
\caption{
\label{fig:pole-zero}
Pole-zero plot of the reflection amplitude $r(z)$ formulated in Eq.~(\ref{eq:reflection-coeff-roots}). Perylene terminus is illustrated as an example with $Q_{\rm A}=7$ and $Q_{\rm T}=3$. The poles marked as ``$\times$" are attributed to the bound states as well as the zeros (``$\circ$") in blue, whereas the zeros are related to the resonant states (in red) and unphysical solutions (in green).
}
\end{figure}

The structure of resonances of $r(z)$ is schematically shown in Fig.~\ref{fig:pole-zero}. We obtain $3$ poles on the real axis corresponding to the $3$ bound states, as well as the $3$ zeros (in blue) associated with the poles; a ``trivial'' zero at $z=0$ and another in the vicinity (both in green); and $Q_{{\rm T}}=4$ pairs of resonances (in red) close to the circle that represent the $4$ resonant states, shown as scattering phase kinks in Fig.~\ref{fig:perylene-linear}c. As a result, the sharp resonances have been accurately reproduced in terms of $k$ (Fig.~\ref{fig:perylene-linear}c). In addition, energies of the bound states in a semi-infinite chain located outside of the exciton band, can be easily found using Eq.~(\ref{eq:spectrum}). Although the nearest-neighbor lattice models are less accurate in terms of exciton spectrum \cite{ES-lattice}, the bound state energies have been qualitatively well reproduced (Table.~\ref{tab:bound_states}).

\begin{figure*}
  \includegraphics[scale=0.85, trim={2cm 13cm 2cm 2cm},clip]{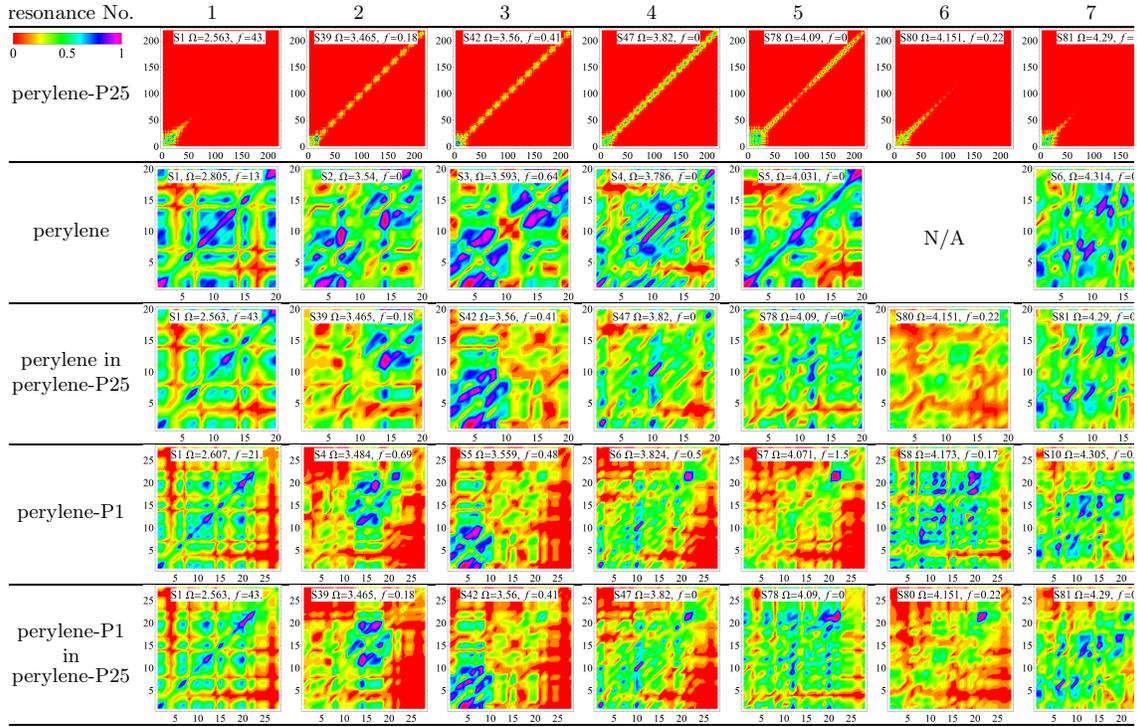}
\caption{Electronic excitations, related to bound and resonant states, given by the contour plots of the transition density matrices from the ground state to excited states of perylene-substituted linear PA molecule with 25 repeat units (denoted by perylene-P$25$), perylene, and phenylethynyl perylene (perylene-P$1$). The axis labels represent indices of carbon atoms starting from perylene (1 to 20) and along the polymer chain. The inset of each plot shows the electronic mode number, the excitation energy $\Omega$ and the oscillator strength $f$.
\label{fig:contour-plots}
}
\end{figure*}

Thus far, we have described the electronic excitations in perylene-attached PA chain (independent of the length of the polymer) by characterizing the analytical and topological properties of the corresponding exciton scattering matrix. The aforementioned tight-binding model, which relies upon the resonances between the states of perylene and the exciton band of the semi-infinite PA chain, is constructed by inspecting electronic excited states in perylene-substituted PA molecules. Stated differently, the tight-binding graph representing the perylene substitution (Fig.~\ref{fig:perylene-linear}b) is determined by the numbers of the bound and resonant excitations that we observed in the substituted linear molecules, without really performing QC analysis on the molecule of perylene. Although the above tight-binding model excellently characterizes electronic excitations in such substituted molecules, the QC calculations found only 6 excitations (not 7) in perylene. Of these, only 5 of 6 states (states 1, 2, 3, 4, and 7 in Fig.~\ref{fig:contour-plots}) are considered to be relevant to the resonances in perylene-substituted PA molecules (hereafter referred to as perylene-P$L$, $L$ being the length of the attached linear segment). This observation indicates that the effect of perylene substitution on the electronic excitations can not be exactly interpreted as just resonances between states of perylene and of the chain; the analytical structure of the scattering coefficients can have substantial differences compared to the one predicted based on just resonances. Such modification of $r(z)$ reflects the fact that perylene is ``chemically" bonded rather than merely ``physically" resonated with the linear segment.


Indeed, by careful examination of the transition density matrices of bound and resonant states, it is found that all these excitations continuously extend into the first repeat unit of the attached PA segment. In particular, states 5 and 6 are highly localized on the first repeat unit of the chain rather than on the perylene (see Fig.~\ref{fig:contour-plots}), and do not match any excitation in the isolated perylene. Comparing excitation structures between resonant states in perylene-PA molecules and phenylethynyl perylene (
denoted by perylene-$P1$), we found that 7 out of 8 states of perylene-$P1$ are attributed to 7 resonances in the perylene-PA molecules, whereas the remaining state of perylene-$P1$ is just a standing wave that resides on the triple bond of the phenylacetylene, which disappears in elongated linear chains. Regarding chemical substitution, attaching perylene to a PA molecule creates a ``larger room" for electronic excitations on the first phenylacetylene repeat unit. Accordingly, the first repeat unit of the PA segment is more like a part of the terminus rather than of the linear chain itself in terms of electronic excitations. Regarding tight-binding model representation, such detailed analysis on excited-state electronic structures is consequently followed by a different tight-binding morphology, i.e., the terminus being described by 8 lattice sites and 1 less site in the chain. Stated equivalently, more ``chemically'' exact tight-binding model of the substituent could be built upon the substitution chemistry. It is worth to mention that altering the structure of the lattice model, with respect to different choices of the terminus, will not affect its outcome, since the relevant tight-binding parameters are adjusted by fitting the unique exciton scattering properties regardless of what to be included in the terminus.

In conclusion, we demonstrated that the analytical and topological structure of the reflection coefficient $r(z)$, which can be conveniently analyzed using tight-binding (lattice) models, provides complete characterization of effects of terminal chemical substitutions in conjugated molecules on electronic excited-state structures.
This approach can be applied to any chemical substitution in conjugated polymers, which can be treated as molecular vertex and, thus, characterized by exciton scattering matrix within the ES approach. By inspecting the structure of excited states, an effective tight-binding model has been formulated to incorporate the coupling between molecular substituent and the attached molecule. The tight-binding graph representing the substituent provides an analytical expression for the associated scattering matrix. By comparing the scattering matrix of the tight-binding model with the ES counterpart, we parameterize the tight-binding model and completely obtain the analytical property of the exciton scattering. Conducting the analytic continuation of the scattering amplitude to complex values of the exciton quasimomentum, the modifications of electronic excitations are distinguished in terms of just positions of poles and zeros of the scattering amplitude that provide sufficient information on appearing bound and resonant states, respectively. Furthermore, delicate description of the interaction between chemical substituent and conjugated polymer have been attained by inspecting detailed excited-state electronic structures. As a consequence, corresponding tight-binding model can be built in a way not only of ``phenomenological'' exactness but also of ``chemical'' consistency.

Starting from quantum-chemical data, processed by the ES method as a bridge, a chemical substitution on a conjugated polymer, in terms of electronic excitations, can be straightforwardly represented by properly constructed tight-binding model. In return, the analytical property of exciton scattering, that adequately describes how electronic excitations in conjugated molecules are affected by the chemical substitution, can be effectively characterized. In other words, the chemical substitution effect on electronic states is fully determined by the excited-state properties of the substituent and the polymer, as well as their couplings, which have been characterized as tight-binding parameters and can provide quick and intuitive guidance for applying chemical substitutions in molecular design of organic semiconductors with desired optoelectronic properties. The described interactions between the substituent and the polymer chain play an important role in dynamical processes involving exciton-phonon couplings, e.g., incoherent energy transfer and charge transportation. Taking into account the simplicity of tight-binding models and their nature resting on the resonance between the substituent and the chain, the ES analysis and tight-binding representations can be feasibly applied to photoinduced dynamics in conjugated macromolecular systems.

This material is based upon work supported by the
National Science Foundation under Grant No. CHE-
1111350. We acknowledge support of the U.S. Department
of Energy through the Los Alamos National Laboratory
(LANL) LDRD Program. LANL is operated by
Los Alamos National Security, LLC, for the National Nuclear
Security Administration of the U.S. Department of
Energy under contract DE-AC52-06NA25396. We acknowledge
support of Center for Integrated Nanotechnology
(CINT) and Center for Nonlinear Studies (CNLS) at
LANL.


\begin{thebibliography}{10}

\bibitem[Landau(1949)]{Landau:superfluidity}
Landau,~L. On the Theory of Superfluidity. \emph{Phys. Rev.} \textbf{1949},
  \emph{75}, 884--885\relax

\bibitem[Tsuneto(1999)]{superfluidity}
Tsuneto,~T. \emph{Superconductivity and Superfluidity}; Cambridge University
  Press, 1999\relax

\bibitem{Kokkelmans:RSuperfluidity}
Kokkelmans,~S. J. J. M.~F.; Milstein,~J.~N.; Chiofalo,~M.~L.; Walser,~R.;
  Holland,~M.~J. Resonance Superfluidity: Renormalization of Resonance
  Scattering Theory. \emph{Phys. Rev. A} \textbf{2002}, \emph{65}, 053617\relax

\bibitem[Landau(1957)]{Landau:Fermi-liquid1957}
Landau,~L. Theory of Fermi-liquids. \emph{Sov. Phys. JETP} \textbf{1957},
  \emph{3}, 920--925\relax

\bibitem[Baym and Pethick(1991)Baym, and Pethick]{Fermi-liquid}
Baym,~G.; Pethick,~C. \emph{Landau Fermi-Liquid Theory}; Wiley: New York,
  1991\relax

\bibitem[Zhang and Zhang(1996)Zhang, and Zhang]{Zhang:GasPhaseReactions}
Zhang,~D.~H.; Zhang,~J. Z.~H. In \emph{Dynamics of Molecules and Chemical
  Reactions}; Wyatt,~R.~E., Zhang,~J. Z.~H., Eds.; CRC Press, 1996; Chapter
  6\relax

\bibitem[Rebentrost and Lester(1977)Rebentrost, and Lester]{Rebentrost:JCP1977}
Rebentrost,~F.; Lester,~J.,~W.~A. Nonadiabatic Effects in the Collision of
  F($^2 P$) with H2($^1 \Sigma _g ^+$). III. Scattering Theory and
  Coupled-Channel Computations. \emph{J. Chem. Phys.} \textbf{1977}, \emph{67},
  3367--3375\relax

\bibitem[Cooper(1983)]{FunctionalAnalysis}
Cooper,~J. In \emph{Functional Analysis, Holomorphy, and Approximation Theory};
  Zapata,~G.~I., Ed.; CRC Press, 1983; Vol.~83\relax

\bibitem{ES-naturephys}
Wu,~C.; Malinin,~S.~V.; Tretiak,~S.; Chernyak,~V.~Y. Exciton Scattering and
  Localization in Branched Dendrimeric Structures. \emph{Nature Phys.}
  \textbf{2006}, \emph{2}, 631 -- 635\relax

\bibitem{ES-prl}
Wu,~C.; Malinin,~S.~V.; Tretiak,~S.; Chernyak,~V.~Y. Multiscale Modeling of
  Electronic Excitations in Branched Conjugated Molecules Using an Exciton
  Scattering Approach. \emph{Phys. Rev. Lett.} \textbf{2008}, \emph{100},
  057405\relax

\bibitem{ES1}
Wu,~C.; Malinin,~S.~V.; Tretiak,~S.; Chernyak,~V.~Y. Exciton Scattering
  Approach for Branched Conjugated Molecules and Complexes. I. Formalism.
  \emph{J. Chem. Phys.} \textbf{2008}, \emph{129}, 174111\relax

\bibitem{ES2}
Wu,~C.; Malinin,~S.~V.; Tretiak,~S.; Chernyak,~V.~Y. Exciton Scattering
  Approach for Branched Conjugated Molecules and Complexes. II. Extraction of
  the Exciton Scattering Parameters from Quantum-Chemical Calculations.
  \emph{J. Chem. Phys.} \textbf{2008}, \emph{129}, 174112\relax

\bibitem{ES3}
Wu,~C.; Malinin,~S.~V.; Tretiak,~S.; Chernyak,~V.~Y. Exciton Scattering
  Approach for Branched Conjugated Molecules and Complexes. III. Applications.
  \emph{J. Chem. Phys.} \textbf{2008}, \emph{129}, 174113\relax

\bibitem{ES-dipole}
Li,~H.; Malinin,~S.~V.; Tretiak,~S.; Chernyak,~V.~Y. Exciton Scattering
  Approach for Branched Conjugated Molecules and Complexes. IV. Transition
  Dipoles and Optical Spectra. \emph{J. Chem. Phys.} \textbf{2010}, \emph{132},
  124103\relax

\bibitem{ES-DA}
Li,~H.; Wu,~C.; Malinin,~S.~V.; Tretiak,~S.; Chernyak,~V.~Y. Excited States of
  Donor and Acceptor Substituted Conjugated Oligomers: A Perspective from the
  Exciton Scattering Approach. \emph{J. Phys. Chem. Lett.} \textbf{2010},
  \emph{1}, 3396--3400\relax

\bibitem{ES-NAO}
Li,~H.; Chernyak,~V.~Y.; Tretiak,~S. Natural Atomic Orbital Representation for
  Optical Spectra Calculations in the Exciton Scattering Approach. \emph{J.
  Phys. Chem. Lett.} \textbf{2012}, \emph{3}, 3734--3739\relax


\bibitem[Heeger(2001)]{Heeger:NobelLecture}
Heeger,~A.~J. Nobel Lecture: Semiconducting and Metallic Polymers: The Fourth
  Generation of Polymeric Materials. \emph{Rev. Mod. Phys.} \textbf{2001},
  \emph{73}, 681--700\relax

\bibitem{Heeger:RevModPhys1988}
Heeger,~A.~J.; Kivelson,~S.; Schrieffer,~J.~R.; Su,~W.~P. Solitons in
  Conducting Polymers. \emph{Rev. Mod. Phys.} \textbf{1988}, \emph{60},
  781--850\relax

\bibitem{BredasJL:Excesc}
Br\'{e}das,~J.~L.; Cornil,~J.; Beljonne,~D.; dos Santos,~D.~A.; Shuai,~Z.
  Excited-State Electronic Structure of Conjugated Oligomers and Polymers: A
  Quantum-Chemical Approach to Optical Phenomena. \emph{Acc. Chem. Res.}
  \textbf{1999}, \emph{32}, 267--276\relax

\bibitem[Heeger(2010)]{heeger2010semiconducting}
Heeger,~A.~J. Semiconducting Polymers: the Third Generation. \emph{Chem. Soc.
  Rev.} \textbf{2010}, \emph{39}, 2354--2371\relax

\bibitem[Dreuw and Head-Gordon(2005)Dreuw, and Head-Gordon]{DreuwA:Sinaim}
Dreuw,~A.; Head-Gordon,~M. Single-Reference Ab Initio Methods for the
  Calculation of Excited States of Large Molecules. \emph{Chem. Rev.}
  \textbf{2005}, \emph{105}, 4009 -- 4037\relax

\bibitem{Mullen:ChemRev1999}
Berresheim,~A.~J.; Muller,~M.; Mullen,~K. Polyphenylene Nanostructures.
  \emph{Chem. Rev.} \textbf{1999}, \emph{99}, 1747--1786\relax

\bibitem{KopelmanR:Speeel}
Kopelman,~R.; Shortreed,~M.; Shi,~Z.~Y.; Tan,~W.~H.; Xu,~Z.~F.; Moore,~J.~S.;
  Bar~Haim,~A.; Klafter,~J. Spectroscopic Evidence for Excitonic Localization
  in Fractal Antenna Supermolecules. \emph{Phys. Rev. Lett.} \textbf{1997},
  \emph{78}, 1239--1242\relax

\bibitem{PercecV:Selshd}
Percec,~V.; Glodde,~M.; Bera,~T.~K.; Miura,~Y.; Shiyanovskaya,~I.;
  Singer,~K.~D.; Balagurusamy,~V. S.~K.; Heiney,~P.~A.; Schnell,~I.; Rapp,~A.
  et~al.  Self-Organization of Supramolecular Helical Dendrimers into
  Complex Electronic Materials. \emph{Nature} \textbf{2002}, \emph{419}, 384 --
  387\relax

\bibitem[Bunz(2005)]{Bunz:AdvPolymSci05}
Bunz,~U. H.~F. Poly(arylene etynylene)s: From Synthesis to Application.
  \emph{Adv. Polym. Sci.} \textbf{2005}, \emph{177}, 1--52\relax

\bibitem{ahn:jcp06}
Ahn,~T.~S.; Thompson,~A.~L.; Bharathi,~P.; Muller,~A.; Bardeen,~C.~J.
  Light-Harvesting in Carbonyl-Terminated Phenylacetylene Dendrimers: The Role
  of Delocalized Excited States and the Scaling of Light-Harvesting Efficiency
  with Dendrimer Size. \emph{J. Phys. Chem. B} \textbf{2006}, \emph{110},
  19810--19819\relax

\bibitem{Moore:awie06}
Zhang,~W.; Moore,~J.~S. Shape-Persistent Macrocycles: Structures and Synthetic
  Approaches from Arylene and Ethynylene Building Blocks. \emph{Angew. Chem.
  Int. Ed.} \textbf{2006}, \emph{45}, 4416--4439\relax

\bibitem{HaleyMM:ChemRev06}
Spitler,~E.~L.; Johnson,~C.~A.; Haley,~M.~M. Renaissance of Annulene Chemistry.
  \emph{Chem. Rev.} \textbf{2006}, \emph{106}, 5344--5386\relax

\bibitem{ES-lattice}
Li,~H.; Malinin,~S.~V.; Tretiak,~S.; Chernyak,~V.~Y. Effective Tight-Binding
  Models for Excitons in Branched Conjugated Molecules. \emph{J. Chem. Phys.}
  \textbf{2013}, \emph{139}, 064109\relax

\bibitem{MinamiT:FreHdn}
Minami,~T.; Tretiak,~S.; Chernyak,~V.; Mukamel,~S. Frenkel-Exciton Hamiltonian
  for Dendrimeric Nanostar. \emph{J. Lum.} \textbf{2000}, \emph{87-9},
  115--118\relax

\bibitem{fidder:7880}
Fidder,~H.; Knoester,~J.; Wiersma,~D.~A. Optical Properties of Disordered
  Molecular Aggregates: A Numerical Study. \emph{J. Chem. Phys.} \textbf{1991},
  \emph{95}, 7880--7890\relax

\bibitem{ES-symmetric}
Li,~H.; Wu,~C.; Malinin,~S.~V.; Tretiak,~S.; Chernyak,~V.~Y. Exciton Scattering
  on Symmetric Branching Centers in Conjugated Molecules. \emph{J. Phys. Chem.
  B} \textbf{2011}, \emph{115}, 5465--5475\relax


\bibitem[Cook(1998)]{cook}
Cook,~D.~B. \emph{Handbook of Computational Quantum Chemistry}; Oxford
  University Press: New York, 1998\relax

\bibitem{g09}
Frisch,~M.~J.; Trucks,~G.~W.; Schlegel,~H.~B.; Scuseria,~G.~E.; Robb,~M.~A.;
  Cheeseman,~J.~R.; Scalmani,~G.; Barone,~V.; Mennucci,~B.; Petersson,~G.~A.
  et~al.  Gaussian~09 {R}evision {A}.01. {G}aussian Inc. Wallingford CT
  2009\relax

\bibitem{ChernyakV:KryatH}
Chernyak,~V.; Schulz,~M.~F.; Mukamel,~S.; Tretiak,~S.; Tsiper,~E.~V.
  Krylov-Space Algorithms for Time-Dependent Hartree-Fock and Density
  Functional Computations. \emph{J. Chem. Phys.} \textbf{2000}, \emph{113},
  36--43\relax

\bibitem{TretiakS:Recdam}
Tretiak,~S.; Chernyak,~V.; Mukamel,~S. Recursive Density-Matrix-Spectral-Moment
  Algorithm for Molecular Nonlinear Polarizabilities. \emph{J. Chem. Phys.}
  \textbf{1996}, \emph{105}, 8914--8928\relax

\bibitem[Tretiak and Mukamel(2002)Tretiak, and Mukamel]{TretiakS:Denmas}
Tretiak,~S.; Mukamel,~S. Density Matrix Analysis and Simulation of Electronic
  Excitations in Conjugated and Aggregated Molecules. \emph{Chem. Rev.}
  \textbf{2002}, \emph{102}, 3171--3212\relax

\bibitem[Zerner(1996)]{zindo}
Zerner,~M.~C. \emph{ZINDO, A Semiempirical Quantum Chemistry Program}; Quantum
  Theory Project, University of Florida: Gainesville, FL, 1996\relax

\end{thebibliography}


\newpage

\begin{table}
\begin{center}
\begin{tabular}[b]{c c c c c c c c}
\hline
$\alpha$&1&2&3&4&5&6&7\\
\hline
$\omega_\alpha$&2.716&3.448&3.551&3.713&3.951&4.063&4.209\\
$J_\alpha$&$-0.281$&$-0.057$&$-0.047$&$-0.238$&$-0.152$&$-0.032$&$-0.029$\\
\hline
\end{tabular}
\end{center}
\caption{Tight-binding parameters (in eV) of perylene terminus in PA molecules: on-site energy $\omega_\alpha$ and hopping constant $J_\alpha$.
\label{tab:parameters}
}
\end{table}

\begin{table}
\begin{center}
\begin{tabular}[b]{c c c c}
\hline
bound state&1&2&3\\
\hline
tight-binding model& $-0.312$ & 0.028 & 0.143 \\
quantum chemistry& $-0.263$ & 0.035 & 0.175 \\
\hline
\end{tabular}
\end{center}
\caption{Bound state energies (in eV) relative to nearby exciton band edge predicted by the nearest-neighbor tight-binding model and from QC computations. Negative value indicates the state below the exciton band, whereas positive values correspond to states above the band.
\label{tab:bound_states}
}
\end{table}

\end{document}